\documentclass[twocolumn,twoside,slac_two]{revtex4}
\usepackage{graphicx}
\usepackage{fancyhdr}
\input{babarsym.tex}

\def\ze#1   {\ensuremath{\zeta_{#1}}\xspace}

\def\pepii    {PEP-II\xspace}

\def\superb   {Super\!\!\;$B$\xspace}

\def\CO2  {$\mathrm{CO}_2$\xspace}

\begin{document}

\title{Charm Physics Opportunities at a Super Flavor Factory}

%

\author{D. Asner}
\affiliation{Carleton University, Ottawa, Canada}

\begin{abstract}
The primary physics goals of a high luminosity $e^+e^-$ flavor factory are discussed, including 
the possibilities to perform detailed studies of the CKM mechanism of quark mixing, and 
constrain virtual Higgs and non-standard model particle contributions to the dynamics of rare 
$B_{u,d,s}$ decays. The large samples of $D$ mesons and tau leptons produced at a flavor factory will result in improved sensitivities to rare $D$ processes - mixing, CP violation and rare decays
- and lepton flavor violation searches, respectively. 
Recent developments in accelerator physics have demonstrated the feasibility to build an accelerator that can achieve luminosities of ${\cal O}(10^36)$ cm$^{-2}$ $s^{-1}$ at $\sqrt{s}=10$~GeV.
The capablity to run at $\sqrt{s}=3.770$~GeV with luminosity of 
$10^{35}\,{\rm cm}^{-2} {\rm s}^{-1}$ is included in the initial design.
This report emphasizes the charm physics that can be probed at a Super Flavor Factory.
\end{abstract}

\maketitle

\thispagestyle{fancy}


{\it These proceedings aim to present a brief overview of the Super$B$ effort with a
special emphasis on the charm physics program of such a facility. In the interest of completeness (and time)
some passages from the Super$B$ Conceptual Design Report\cite{Bona:2007qt} are reproduced here.}

\section{Introduction}

Elementary particle physics in the next decade will be focused on the
investigation of the origin of electroweak symmetry breaking and the
search for extensions of the Standard Model (SM) at the TeV scale. The
discovery of New Physics will likely produce a period of excitement
and progress recalling the years following the discovery of the
\jpsi. In this new world, attention will be riveted on the detailed
elucidation of new phenomena uncovered at the LHC; these discoveries
will also provide strong motivation for the construction of the ILC. 
High statistics studies of heavy quarks and leptons will have a
crucial role to play in this new world. 

The  two asymmetric $B$ Factories, \pepii~\cite{ref:PEP-II} and
KEKB\cite{ref:KEKB}, and their companion detectors,
\babar\cite{ref:babar} and Belle\cite{ref:belle}, have over the last
seven years produced a wealth of flavour physics results, subjecting
the quark and lepton sectors of the Standard Model to a series of
stringent tests, all of which have been passed. With the much larger
data sample made possible by a Super~$B$~Factory, qualitatively new
studies will be possible. These studies will provide a uniquely
important source of information about the details of the New Physics
uncovered at hadron colliders in the coming decade. 

The continued detailed studies of heavy quark and
heavy lepton (henceforth {\it heavy flavour}) physics will not only be
pertinent in the next decade; they will be central to understanding
the flavour sector of New Physics phenomena. A Super Flavour Factory
such as Super$B$ will be a partner with LHC, and
eventually, ILC, experiments, in ascertaining exactly {\it what kind}
of New Physics has been found. The capabilities of Super$B$ in
measuring \CP-violating asymmetries in very rare $b$ and $c$ quark
decays, accessing branching fractions of heavy quark and heavy lepton
decays in processes that are either extremely rare or forbidden in the
Standard Model, and making detailed investigations of complex
kinematic distributions will provide unique and important constraints
in, for example, ascertaining the type of supersymmetry breaking or
the kind of extra dimension model behind the new phenomena that many
expect to be manifest at the LHC. 

The Super$B$ Conceptual Design Report\cite{Bona:2007qt} 
is the founding document of a
nascent international enterprise aimed at the construction of a very
high luminosity asymmetric \epem Flavour Factory.  A possible location
for Super$B$ is the campus of the University of Rome ``Tor Vergata'',
near the INFN National Laboratory of Frascati. 

\begin{figure}[htbp]
 \begin{center}
  \includegraphics[width=0.49\textwidth]{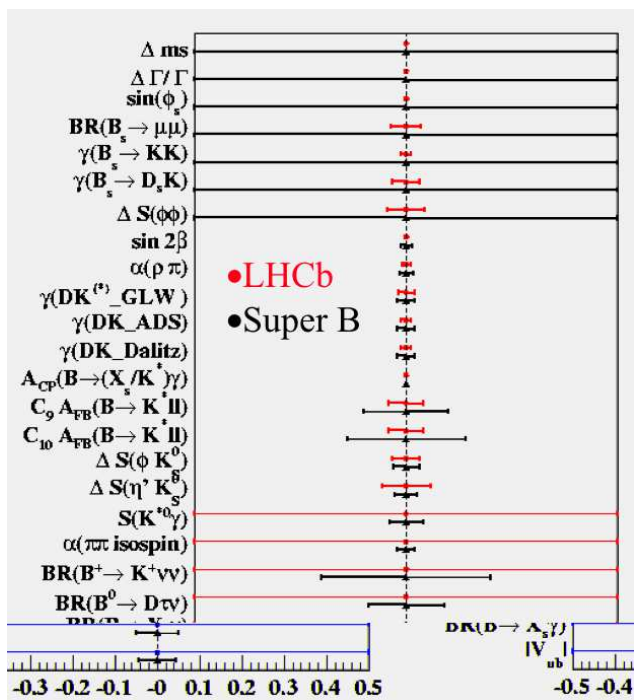}
 \caption{Comparison of Super$B$ with 50 ab$^{-1}$ and and upgraded LHCb 100 fb$^{-1}$. 
Design luminosity for Super$B$ is 15 ab$^{-1}$/year. Design luminosity for LHCb is 2 fb$^{-1}$/year.
This comparison assumes 
that Super$B$ does not integrate luminosity at the \FiveS.
An upgraded LHCb could integrate luminosity at a 10 times greater rate than LHCb.}
  \label{fig:lhcb}
 \end{center}
\end{figure}

The exciting physics program that can be
accomplished with a very large sample of heavy quark and heavy lepton
decays produced in the very clean environment of an \epem collider; 
with a peak luminosity in excess of \hbox{$10^{36}$ cm$^{-2}$ s$^{-1}$} 
at the \FourS resonance is described in Ref.\cite{Bona:2007qt} and summarized below.
This is program complementary to that of an experiment such as \lhcb at a
hadronic machine. The physics reach of \lhcb and Super$B$ in the $b$-sector 
are compared in Figure~\ref{fig:lhcb}. 
Luminosities of $10^{35}\,{\rm cm}^{-2} {\rm s}^{-1}$ at the $\psi(3770)$ are expected.
This report focuses on the charm physics that can be probed both near the \FourS resonance
and near charm production threshold.

The conceptual design of a new type
of \epem collider that produces a nearly two-order-of-magnitude
increase in luminosity over the current generation of asymmetric $B$
Factories is described in Ref.\cite{Bona:2007qt}. 
The key idea is the use of low emittance beams produced in
an accelerator lattice derived from the ILC Damping Ring Design,
together with a new collision region, again with roots in the ILC
final focus design, but with important new concepts developed in this
design effort.  Remarkably, Super$B$ produces this very large
improvement in luminosity with circulating currents and wallplug power
similar to those of the current $B$ Factories. 
The design of an appropriate detector, based on an upgrade of \babar\ as an example, is
also discussed in some detail in Ref.~\cite{Bona:2007qt}. 

\subsection{The Physics Case for Super$B$}

\begin{figure}[htbp]
\begin{center}
  \includegraphics[angle=90, width=0.49\textwidth]{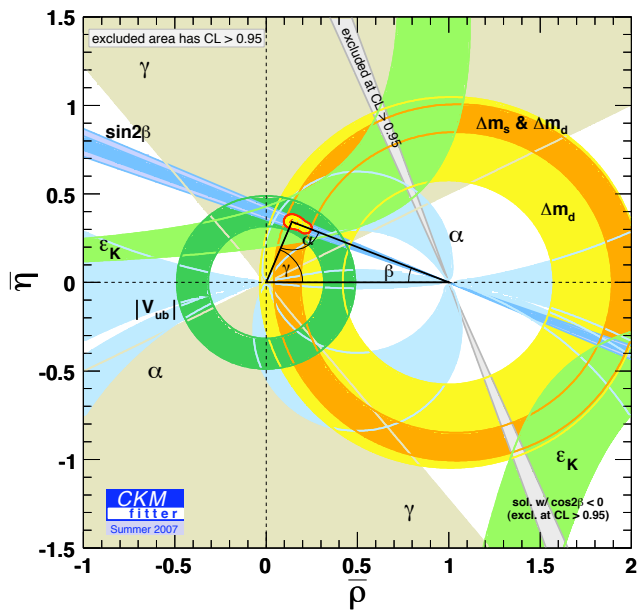}
 \caption{Global fit of the Unitarity Triangle construction as of LeptonPhoton
 2007 conference.} 
  \label{fig:summer07}
 \end{center}
\end{figure}

By measuring mixing-dependent $C\!P$-violating asymmetries in the $B$
meson system for the 
first time, \pepii/\babar\ and KEKB/Belle have shown that the CKM
phase accounts 
for all observed $C\!P$-violating phenomena in $b$ decays.
The Unitarity Triangle construction provides a set of unique
overconstrained precision tests of the self-consistency of the three
generation Standard Model. 
Figure~\ref{fig:summer07} shows the
current status of the Unitarity Triangle construction, incorporating
measurements from \babar\ and Belle, as well as the $B_s$ mixing
measurement of CDF; the addition of \CP asymmetry measurements,
together with the improvement in the precision of \CP-conserving
measurements, has made this uniquely precise set of Standard Model
tests possible.

The fact that the CKM phase has now been shown to be consistent with
all observed \CP-violating phenomena is both a triumph and an
opportunity. In completing the experimentally-verified Standard Model
{\it ansatz} (except, of course, for the Higgs), it intensifies the
mystery of the creation of the baryon-antibaryon asymmetry of the
universe: the observed \CP-violation is too small for the Standard
Model to account for electroweak baryogenesis. 
This intriguing result opens the door to two possibilities: the matter
antimatter asymmetry is produced by another mechanism, such as
leptogenesis, or baryogenesis proceeds through the additional
\CP-violating phases that naturally arise in many extensions of the
Standard Model. 
These extra phases produce measurable effects in the weak decays of
heavy flavour particles. The detailed pattern of these effects, as
well as of rare decay branching fractions and kinematic distributions,
is, in fact, diagnostic of the characteristics of New Physics at or
below the TeV scale, 

By the end of this decade, the two $B$ Factories will have accumulated
a total of \hbox{$\sim 2$ \ab}. Even at this level, most important
measurements pertinent to the Unitarity Triangle construction will
still be statistics limited: an even larger data sample would provide
increasingly stringent tests of three-generation CKM unitarity. There
are two main thrusts here. The first is the substantial remaining
improvement that can still be made in the Unitarity Triangle
construction. 
Here measurements in $B$, $D$ and $\tau$ decay play an important role,
as do improvements in lattice QCD calculations of hadronic matrix
elements. This important physics goal is {\bf NOT}, however, the sole,
or even the primary, motivation for a \sbf. The precision of our
knowledge of the Unitarity Triangle will perforce improve to the limit
allowed by theoretical uncertainties as we pursue the primary goal:
improving the precision of the measurement of \CP\ asymmetries, rare
decay branching fractions, and rare decay kinematic distributions in
penguin-dominated $b\rightarrow s$ transitions, to a level where there
is substantial sensitivity to New Physics effects. 
This requires data samples substantially larger than the
current $B$ Factories will provide. Some of these measurements are
accessible at the LHC~\cite{ref:LHCb},
but the most promising approach to this
physics is Super$B$, a very high luminosity \abf, which is also, of
course, a Super Flavour Factory, providing large samples of $b$ and
$c$ quark and $\tau$ lepton decays. 

Super$B$, having an initial luminosity of $10^{36}$ cm$^{-2}$s$^{-1}$,
will collect 15 ab$^{-1}$ in a New Snowmass
Year~\cite{ref:snowmassyear}, or 75 ab$^{-1}$ in five years. 
A data sample this large will make the Unitarity Triangle tests, in
their manifold versions, the ultimate precision test of the flavour
sector of the Standard Model, and open up the world of New Physics
effects in very rare $B$, $D$, and $\tau$ decays 

A primary tool for isolating new physics is the time-dependent
\CP\ asymmetry in decay channels that
proceed through penguin diagrams, such as
the $b\rightarrow s\bar{s}s$ processes $B^0_d \rightarrow \phi K^0$ and $B^0_d \rightarrow (K\bar{K})_{C\!P} K^0$
or similar transitions such as
$B^0_d \rightarrow \eta 'K^0$,  $B^0_d \rightarrow f_0K^0$,  $B^0_d \rightarrow \pi^0 K^0$,
$B^0_d \rightarrow \rho^0 K^0$, $B^0_d \rightarrow \omega K^0$, and $B^0_d \rightarrow \pi^0\pi^0 K^0$.
The dominant contribution to these decays is the combination of CKM elements $V_{tb}V_{ts}^{*}$; these amplitudes have the
same phase as
the charmonium channels $b \rightarrow c\bar{c}s$,
up to a small phase shift of $V_{ts}$ with respect to $V_{cb}$. New heavy particles contribute new loop amplitudes, with new phases that
can contribute to the 
$C\!P$ asymmetry and the $S$ coefficient of the time-dependent analysis,
so that the measured \CP violation parameter
could be substantially different from $\sin 2\beta$.

Physics beyond the Standard Model can affect rare $B$ decay modes, through observables such as branching fractions, $C\!P$-violating asymmetries and kinematic distributions.
These decays do not typically occur at tree level,
and thus their rates are strongly suppressed in the Standard Model. Substantial enhancements in the
rates and/or variations in angular distributions of final state particles could result from the presence of new
heavy particles in loop diagrams, resulting in clear evidence of New Physics. Moreover, because the pattern
of observable effects is highly model-dependent, measurements of several rare decay modes can provide
information regarding the source of the New Physics. An extended run at the \FiveS is also contemplated; such a run would yield a wealth of interesting new $B_s^0$ decay results.

The Super$B$ data sample will also contain unprecedented numbers of charm quark and $\tau$ lepton decays. This data is also of great interest, both for its capacity to improve the precision of existing measurements and for its sensitivity to New Physics. This interest extends beyond weak decays; the detailed exploration of new charmonium states is also an important objective. Limits on rare $\tau$ decays, particularly lepton-flavour-violating decays, already provide important constraints on New Physics models. Super$B$ may have the sensitivity to actually observe such decays. The accelerator design will allow for longitudinal polarization of the $e^-$ beam, making possible uniquely sensitive searches for a $\tau$ electric dipole moment, as well as for \CP-violating $\tau$ decays.

Some measurements in charm and $\tau$ physics are best done near threshold. Super$B$ also has the capability of running in the 4 \gev region. Short runs at specific center-of-mass energies in this region, representing perhaps 10\% of data taking time, would produce data samples substantially larger than those currently envisioned to exist in the next decade. The many advantages of 
analysis at threshold are enumerated in Section~\ref{sec:thresh}

\subsection{The Super$B$ Design}

Given the strong physics motivation, there has been a great deal of activity over the past few years aimed at designing an $e^+e^-$ $B$ Factory that can produce samples of $B$ mesons 50 to 100 times larger than will exist when the current $B$ Factory programs end. Several approaches were tried before the design\cite{Bona:2007qt} described briefly here was developed.

Upgrades of \pepii~\cite{ref:SuperPEPII} and
KEKB~\cite{ref:SuperKEKB} to Super $B$ Factories that accomplish this goal have been proposed
at SLAC and at KEK. These machines are extrapolations of the existing $B$ Factories, with higher currents, more bunches, and smaller $\beta$ functions (1.5 to 3 mm). They also use a great deal of power ($\ge 100$ MW), and the high currents (as much as 10A) pose significant challenges for detectors. To minimize the substantial wallplug power, the SuperPEP-II design doubled the current RF frequency, to 958 MHz. In the case of SuperKEKB, a factor of two increase in luminosity is assumed for the use of crab
crossing, which will soon be tested at KEKB.

SLAC has no current plans for an on-site accelerator-based
high energy physics program, so the SuperPEP-II proposal
is moribund. As of this writing, no decision has been made on SuperKEKB.
In the interim, the problematic power consumption and background issues associated with the SLAC and KEK-based Super $B$ Factory designs stimulated a new approach, using low emittance beams, to constructing a Super $B$ Factory with a luminosity of $10^{36}$, but with reduced power consumption~\cite{ref:linac1}.

The current machine concept, which has roots in ILC R\&D: a very low emittance storage ring, based on the ILC damping ring minimum emittance growth lattice and final focus, that incorporates several novel accelerator concepts and appears capable of meeting all design criteria, while reducing the power consumption, which dominates the operating costs of the facility, to a level similar to that of the current $B$ Factories. Due to similarities in the design of the low emittance rings and the final focus, operation of Super$B$ can serve as a system test for these linear collider components

By utilizing concepts developed for the ILC damping rings and final focus in the design of the Super$B$ collider, it is possible to produce a two-order-of-magnitude increase in luminosity with beam currents that are comparable to those in the existing asymmetric $B$ Factories.  Background rates and radiation levels associated with the circulating currents are comparable to current values; luminosity-related backgrounds such as those due to radiative Bhabhas, increase substantially. With careful design of the interaction region, including appropriate local shielding, and straightforward revisions of detector components, upgraded detectors based on \babar\ or Belle are a good match to the machine environment:  in this discussion, we use \babar\ as a specific example.  Required detector upgrades include: reduction of the radius of the beam pipe, allowing a first measurement of track position closer to the vertex and improving the vertex resolution (this allows the energy asymmetry of the collider  to be reduced to 7 on 4 GeV); replacement of the drift chamber, as the current chamber will have exceeded its design lifetime; replacement of the endcap calorimeter, with faster crystals having a smaller Moli{\`e}re radius, since there is a large increase in Bhabha electrons in this region.

The Super$B$ design has been undertaken subject to two important
constraints: 1) the lattice is closely related to the ILC
  Damping Ring lattice, and 2) as many \pepii components as possible
have been incorporated into the design. A large number of \pepii
components can, in fact, be reused: The majority of the HER and LER
magnets, the magnet power supplies, the RF system, the digital
feedback system, and many vacuum components. This will reduce the cost
and engineering effort needed to bring the project to fruition. 

The Super$B$ concept is a breakthrough in collider design. The invention of the ``crabbed waist'' final focus can, in fact, have impact even on the current generation of colliders. A test of the crabbed waist concept is planned to take place at Frascati in 2007; a positive result of this test would be an important milestone as the Super$B$ design progresses. The low emittance lattice, fundamental as well to the ILC damping ring design, allow high luminosity with modest power consumption and demands on the detector.

Super$B$ appears to be the most promising approach to
producing the very high luminosity \abf\ that is required to observe
and explore the contributions of physics beyond the Standard Model
to heavy quark and $\tau$ decays.


\graphicspath{{./}{../}{figures/}{Physics/}{Physics/figures/}}


\def\superb{\mbox{\normalfont Super$B$\xspace}}
\def\sbf{\mbox{\normalfont \superb\ Factory}\xspace}
\def\sff{\mbox{\normalfont Super Flavour Factory}\xspace}
\def\babar{\mbox{\slshape B\kern-0.1em{\smaller A}\kern-0.1em
    B\kern-0.1em{\smaller A\kern-0.2em R}}\xspace}
\def\belle{\mbox{\normalfont Belle}\xspace}

\def\etal{{\it et al.}}
\def\ie{{\it i.e.}}
\def\eg{{\it e.g.}}
\def\etc{{\it etc.}}
\def\cf{{\it cf.}}
\def\vs{{\it vs.}}

\def\B{{\ensuremath{{\cal B}}\xspace}}

\newcommand{\subsubsubsection}[1]{\vspace{2ex}\par\noindent {\bf\boldmath\em #1} \vspace{2ex}\par}
\newcommand{\mysection}[1]{\section[#1]{\boldmath #1}}
\newcommand{\mysubsection}[1]{\subsection[#1]{\boldmath #1}}
\newcommand{\mysubsubsection}[1]{\subsubsection[#1]{\boldmath #1}}
\newcommand{\mysubsubsubsection}[1]{\subsubsubsection{\boldmath #1}}


\renewcommand{\Re}{{\rm Re}\,}
\renewcommand{\Im}{{\rm Im}\,}

\newcommand{\GeV}{\,\mbox{GeV}}
\newcommand{\MeV}{\,\mbox{MeV}}
\newcommand{\aver}[1]{\langle #1\rangle}
\newcommand{\matel}[3]{\langle #1|#2|#3\rangle}
\newcommand{\state}[1]{|#1\rangle}
\newcommand{\ve}[1]{\vec{\bf #1}}

\newcommand{\note}[1]{\marginpar{\hspace*{-2mm}\tiny #1}}


\def\mtiny{\vrule width 0pt}
\def\mrm#1{\mathrm{#1}}
\def\DZ{\relax\ifmmode{D^0}\else{$\mrm{D}^{\mrm{0}}$}\fi}
\def\KZ{\relax\ifmmode{K^0}\else{$\mrm{K}^{\mrm{0}}$}\fi}
\def\BZ{\relax\ifmmode{B^0}\else{$\mrm{B}^{\mrm{0}}$}\fi}
\def\BZS{\relax\ifmmode{B_s^0}\else{$\mrm{B_s}^{\mrm{0}}$}\fi}
\def\DZB{\relax\ifmmode{\overline{D}\mtiny^0}
        \else{$\overline{\mrm{D}}\mtiny^{\mrm{0}}$}\fi}
\def\KZB{\relax\ifmmode{\overline{K}\mtiny^0}
        \else{$\overline{\mrm{K}}\mtiny^{\mrm{0}}$}\fi}
\def\BZB{\relax\ifmmode{\overline{B}\mtiny^0}
        \else{$\overline{\mrm{B}}\mtiny^{\mrm{0}}$}\fi}
\def\BZBS{\relax\ifmmode{\overline{B_s}\mtiny^0}
        \else{$\overline{\mrm{B_s}}\mtiny^{\mrm{0}}$}\fi}

\newcommand{\MSbar}{\hbox{$\overline{MS}$\ }}
\newcommand{\ket}[1]{\left| {#1} \right\rangle}
\newcommand{\bra}[1]{\left\langle {#1}\right|}
\newcommand{\braket}[2]{\left\langle {#1} \right| \left. {#2} \right\rangle}
\newcommand{\too}{\mathop{\ \longrightarrow}\ }
\newcommand\eqn[1]{\label{eq:#1}} 
\newcommand\eq[1]{eq.~(\ref{eq:#1})} 
\newcommand\tab[1]{{\footnotesize {\bf Table}~[{\bf\ref{#1}}]}} 
\newcommand\sect[1]{sec.~\ref{sec:#1}} 
\newcommand{\Det}{\mathop{\rm Det}}
\newcommand{\bfn}{{\bf n}}
\newcommand{\bfr}{{\bf r}}
\newcommand{\bfw}{{\bf w}}
\newcommand{\bfm}{{\bf m}}
\newcommand{\bfe}{{\bf  e}}
\newcommand{\xh}{{\bf \hat x}}
\newcommand{\yh}{{\bf \hat y}}
\newcommand{\zh}{{\bf \hat z\,}}
\newcommand{\Tr}{\mathop{\rm Tr}}
\newcommand{\sla}[1]%
        {\kern .25em\raise.18ex\hbox{$/$}\kern-.75em #1}
\newcommand{\mybar}[1]%
        {\kern 0.8pt\overline{\kern -0.8pt#1\kern -0.8pt}\kern 0.8pt}
\newcommand{\cs}{charged scalar}
\newcommand{\css}{charged scalars}
\newcommand{\nfcl}{\mbox{$95\%~{\rm CL}$}}
\newcommand{\ncl}{\mbox{$90\%~{\rm CL}$}}
\newcommand{\brsm}{\mbox{${\rm BR}^{SM}$}}
\newcommand{\brmh}{\mbox{${\rm BR}^{MHDM}$}}
\newcommand{\bxtaunu}{\mbox{$B \rightarrow X \tau \nu_{\tau}$}}

\newcommand{\bbr}{$B${\footnotesize $A$}$B${\footnotesize $AR$} }
\newcommand{\ov}[1]{\overline{#1}}

\newcommand{\msbar}{\overline{\rm MS}}
\newcommand{\bea}{\begin{eqnarray}}
\newcommand{\eea}{\end{eqnarray}}
\newcommand{\beq}{\begin{equation}}
\newcommand{\eeq}{\end{equation}}
\newcommand{\kkbar}{K^0-\overline K^0}
\newcommand{\bbbard}{\overline B_d^0-B_d^0}
\newcommand{\bbbars}{\overline B_s^0-B_s^0}
\newcommand{\dmd}{\Delta m_d}
\newcommand{\dms}{\Delta m_s}

\def\simge{\mathrel{\rlap{\raise 0.511ex \hbox{$>$}}{\lower 0.511ex
\hbox{$\sim$}}}}
\def\simle{\mathrel{\rlap{\raise 0.511ex \hbox{$<$}}{\lower 0.511ex
\hbox{$\sim$}}}}







\mysection{Charm Physics at Super$B$}
\label{sec:charm}
It is a truth universally accepted that charm studies played
a seminal role in the evolution and acceptance of the Standard Model.
Yet the continuing importance of this sector is not widely appreciated,
since the Standard Model electroweak phenomenology for charm decays is on the dull side:
the CKM parameters are known, $\DzDzb$ oscillations are slow,
$\CP$ asymmetries are small or absent and loop-driven decays are extremely rare.

Yet on closer examination, a strong case emerges in two respects,
both of which derive from this apparent dullness:
\begin{itemize}
\item
  Detailed and comprehensive analyses of charm transitions
  will continue to provide us with new insights into QCD's
  nonperturbative dynamics, and advance us significantly
  towards establishing theoretical control over them.
  Beyond the intrinsic value of such lessons, they will also
  calibrate our theoretical tools for $B$ studies;
  this will be essential to saturate the discovery potential for
  New Physics in $B$ transitions.
\item
  Charm decays constitute a novel window into New Physics.
\end{itemize}
Lessons from the first item will have an obvious impact on the tasks
listed under the second.
They might actually be of great value even beyond QCD,
if the New Physics anticipated for the TeV scale
is of the strongly interacting variety.

The capabilities of a \sff\ are well matched to these goals.
It allows uniquely clean determinations of CKM parameters,
with six of the nine matrix elements impacted by charm measurements.
New Physics signals can easily exceed Standard Model predictions by considerable factors
such that there will be no ambiguity in interpreting them,
yet they are unlikely to be large;
this again requires the clean environment and huge statistics
that a \sff\ can provide.

A number of other facilities either currently running or
soon to commence operation provide competition in the area of charm physics.
The current $B$ Factory program is expected to produce a sample of about
$10^{10}$ charm hadrons from operation at or near the
$\FourS$ resonance.
The CLEO-$c$ experiment at CESR
is operating in the charm threshold region,
and anticipates collecting a total of $5 \times 10^6$ $\DzDzb$ pairs
and about $7 \times 10^5$ $D_s^{*+}D_s^- + D_s^{+}D_s^{*-}$
through coherent production.
The BESIII experiment at BEPCII expects first $e^+e^-$ collisions in 2008,
and will collect large charmonium samples,
in addition to exceeding the CLEO-$c$ data set in open charm production.
Although there will be no successors to the Fermilab fixed target
charm production experiments,
the LHC will produce copious quantities of charm
(notably, charm physics forms a part of the \lhcb\ physics program);
these are expected to result in very large samples of charmed hadrons
in final states with reconstructible topologies.

Most of the benchmark charm measurements will still be statistics-limited
after the CLEO-$c$, BESIII and $B$ Factory projects,
and many will not be achievable in a hadronic environment.
\superb\ is the ideal machine with which to pursue
these measurements to their ultimate precision.
Operation near the $\FourS$ will provide
enormous samples of charm hadrons,
in a clean environment and with a detector well-suited for charm studies.
The charm physics program would benefit further from the ability
to operate in the threshold region, in order to
exploit the quantum correlations associated with coherent production.
The expected lower luminosity at threshold would be partly compensated by the
higher production cross-section,
resulting in a comparable charm production rate.
To estimate the reach of \superb\ from operation at the charm threshold,
we have assumed a simple dependence of the luminosity
on the center-of-mass energy: ${\cal L}_{\rm peak} \propto s$.
Thus, we expect that \superb\ (which will integrate
$\sim 15 \ {\rm ab}^{-1}$ per year operating at the $\Upsilon(4{\rm S})$)
can accumulate $\sim 150 \ {\rm fb}^{-1}$ per month
when operated at the $\psi(3770)$.

\subsection{Advantages of Threshold Production}
\label{sec:thresh}
The production rate of charm during threshold running at a Super$B$
and $\Upsilon$(4S) running is comparable. Although
the luminosity for charm threshold running is expected to be an order of
magnitude lower, the production cross section is 3 times higher
than at $\sqrt{s}=10.58$~GeV.
Charm threshold data has distinct and powerful advantages over continuum
and $b \to c$ charm production data accumulated above $B$ production threshold.

\noindent{\bf Charm Events at Threshold are Extremely Clean:}
The charged and neutral multiplicites in $\psi(3770)$ events are only 5.0 and
2.4 - approximately 1/2 the multiplicity of continuum charm production at $\sqrt{s}=10.58$~GeV. 

\noindent{\bf Charm Events at Threshold are pure $D\overline D$:}
No additional fragmentation particles are produced.  The same is true for
$\sqrt{s}=4170$~MeV production of $D\overline D^*$, $D^+_sD^-_s$, 
$D^+_sD^{*-}_s$, and for threshold production of $\Lambda_c \bar \Lambda_c$.
This allows use of kinematic constraints, such as total candidate energy
and beam constrained mass, and permits effective use of missing mass methods
and neutrino reconstruction. The crisp definition of the initial state is a
uniquely powerful advantage of threshold production that is absent in
continuum charm prodution.

\noindent{\bf Double Tag Studies are Pristine:}
The pure production of $D\overline D$ states, together with low multiplicity and
large branching ratios characteristic of many $D$ decays permits effective
use of double-tag studies in which one $D$ meson is fully reconstructed and the
rest of the event is examined without bias but with substantial kinematic 
knowledge. The techniques pioneered by Mark III and extended by 
CLEO-c\cite{He:2005bs,Dobbs:2007zt} allow
precise absolute branching fraction determination. Backgrounds under these
conditions are heavily suppressed which minimizes both statistical errors and
systematic uncertainties.

\noindent{\bf Signal/Background is Optimum at Threshold:}
The cross section for the signal $\psi(3770) \to D\overline D$ is 
about 1/2 the 
cross section for the underlying continuum $e^+e^- \to$ hadrons background.
By contrast, for $c\bar c$ production at $\sqrt{s}=10.58$~GeV the signal is
only 1/4 of the total hadronic cross section. 

\noindent{\bf Neutrino Reconstruction:}
The undetected energy and momentum is interpreted as the neutrino four-vector.
For leptonic
and semileptonic charm 
decays the signal is observed in missing mass squared distributions
and for double-tagged events these measurements have low backgrounds. The
missing mass resolution is about one pion mass. For semileptonic decays
the $q^2$ resolution is excellent, about 3 times better than in continuum
charm reconstruction at $\sqrt{s}=10.58$~GeV. Neutrino reconstruction at threshold
is clean.

\noindent{\bf Quantum Coherence:}
The production
of $D$ and $\overline D$ in a coherent $C=-1$ state from $\psi(3770)$ decay is 
of central importance for the subsequent evolution and decay of these 
particles. The same is true for $D\overline D (n)\pi^0 (m) \gamma$ produced
at $\sqrt{s}\sim 4$~GeV where $C=-1$ for even $m$ and $C=+1$ for odd
$m$. The coherence of the two initial state $D$ mesons allows both simple
and sophisticated methods to measure $D\overline D$ mixing parameters, strong
phases, $CP$ eigenstate branching fractions, and $CP$ 
violation\cite{Bigi:1989ah, Asner:2005wf, Gronau:2001nr, Bigi:2000yz, Bianco:2003vb}.

\subsection{Lessons on Strong Dynamics}

Detailed analyses of (semi)leptonic decays of charm hadrons
provide a challenging test bed for validating lattice QCD (LQCD),
which is the only known framework with realistic promise for a truly
quantitative treatment of charm hadrons that can be systematically improved .
Such studies form the core of the ongoing CLEO-$c$
and the nascent BESIII programs;
they are also pursued very profitably at the $B$~Factories.
Central goals are measuring the decay constants $f_{D^+}$ and $f_{D_s}$
and going beyond total rates for semileptonic $D^+$, $D^0$ and $D_s^+$ decays.
 on the Cabibbo allowed and forbidden level by extracting the form factors \etc
 It is essential to analyze lepton spectra
 and perform ``meaningful'' Dalitz plot studies.
 To quantify ``meaningful'' we can compare to analyses on $K_{e4}$ decays.
 With a sample size of 30,000 events as it became available in 1977
 one was able to begin extracting dynamical information.
 Precise measurements are possible now
 with NA48/2 and E685 each having accumulated 400,000 events.
 For charm we are nowhere near that level yet:
 CLEO-$c$ will have about 10,000 semileptonic charm decays --
 comparable to kaon studies in the late 1970s.
 Since for charm the phase space is larger
 (actually a good thing, since it opens up more domains of interest)
 it seems reasonable to aim for sample sizes of $10^6$ events.
 Again, this is well beyond the reach of CLEO-$c$
 and most probably of BESIII as well.
Such high quality studies will greatly improve
our understanding of hadronization and provide an even richer test bed
for LQCD with the lessons to be learned
of crucial importance for extracting $V_{ub}$ from semileptonic $B$ decays.
Our knowledge of charm baryon decays is also rather limited;
\eg, no precision data on absolute
branching ratios or semileptonic decay distributions exist.
CLEO-$c$ will not run above the charm baryon threshold, and BESIII cannot.

\mysubsubsection{Leptonic Charm Studies}
\label{LEPCHARM}

In the Standard Model the leptonic decay width is given by~\cite{Silverman:1988gc}:
\begin{eqnarray}
  \Gamma(D^+ \!\to\! \ell^+\nu)\! & \!= \! & \!
  \frac{G_F^2}{8\pi} f_{D^+}^2 m_{\ell}^2 M_{D^+}
  \!\left( 1 \! - \! \frac{m_{\ell}^2}{M_{D^+}^2} \right)^2\! \left|V_{cd}\right|^2
  \nonumber \\
  \Gamma(D_s^+\! \to\! \ell^+\nu)\! &\! = \!& \!
  \frac{G_F^2}{8\pi} f_{D^+_s}^2 m_{\ell}^2 M_{D^+_s}
  \!\left( 1\! - \!\frac{m_{\ell}^2}{M_{D^+_s}^2}\right)^2\! \left|V_{cs}\right|^2\,.
  \label{eq:equ_rate}
\end{eqnarray}
Taking $|V_{cd}|$ and $|V_{cs}|$ from elsewhere, one uses Eq.(\ref{eq:equ_rate})
to extract $f_{D^+}$ and $f_{D^+_s}$.
The ratio $R_\ell$ of the leptonic decay rates of the $D_s^+$ and the $D^+$
is proportional to $(f_{D^+_s}/f_{D^+})^2$,
for which the lattice calculation is substantially more precise.
A significant deviation from its predicted value would be a clear
sign of New Physics,
probably in the form of a charged Higgs exchange~\cite{Akeroyd:2003jb}.
On the other hand, the ratio of the rates of tauonic and muonic decays
for either $D^+$ or $D^+_s$ is independent of both form factors
and CKM elements, and serves as a useful cross-check in this context.

CLEO-$c$ has published a measurement of
$f_{D^+}$~\cite{Artuso:2005ym,Bonvicini:2004gv,Rubin:2006nt},
and several measurements of
$f_{D_s^+}$~\cite{Pedlar:2007za,Artuso:2007zg,Stone:2007dy}.
These measurements have benefitted from a ``double-tag'' method
uniquely possible at threshold,
where a $D^+_{(s)}D^-_{(s)}$ pair is produced with no extra particles.
The latest results are
\begin{equation}
  f_{D^+} = (222.6 \pm 16.7 \,^{+2.8}_{-3.4})~{\rm MeV}~.
\end{equation}
\begin{equation}
  f_{D_s} = (275 \pm 10 \pm 5) {~\rm MeV}
\end{equation}
\begin{equation}
  f_{D_s^+}/f_{D^+} = 1.24 \pm 0.10 \pm 0.03~.
\end{equation}
\babar has also measured $f_{D_s} = (283 \pm 17 \pm 7 \pm 14){~\rm MeV}$\cite{Aubert:2006sd}.
The central values for $f_{D_s^+}$ and $f_{D_s^+}/f_{D^+}$
are slightly above, but consistent with, the present LQCD calculations.
It is important to note that the desired $1$--$3\%$ accuracy level
has not yet been reached on either the experimental or theoretical side.
While LQCD practitioners expect to reach this level over the next decade,
the experimental precision is likely to fall significantly short
of this goal, even after BESIII.
Since larger statistics will certainly allow reduction
of the systematic errors in the current results,
it is clear that data accumulated by \superb\
from a relatively short run ($\sim 1$ month) at charm threshold
would allow the desired improvement of the experimental precision
(see discussion below, and Table~\ref{tab:statl}).
Validating LQCD on the ${\cal O}(1\%)$ level will have
important consequences for $B_d$ and $B_s$ oscillations,
since it would give us demonstrated confidence in the theoretical
extrapolation to $f_{B_d}$ and $f_{B_s}/f_{B_d}$.

\mysubsubsection{Semileptonic Charm Studies}

In the area of semileptonic decays,
CLEO-$c$ has made the most accurate measurements for
the inclusive $D^0$ and $D^+$
semileptonic branching fractions --
${\cal B}(D^0 \rightarrow X \ell \nu_\ell)= (6.46 \pm 0.17 \pm 0.13)\%$ and
${\cal B}(D^+ \rightarrow X \ell \nu_\ell)=(16.13 \pm 0.20 \pm 0.33)\%$~\cite{Adam:2006nu} --
and expects to do the same for $D_s^+$.
Such data provide important ``engineering input''
for other $D$ and $B$ decay studies.
However, a central goal must be to go beyond the total rates for these decays
and to extract the form factors \etc\
In order to do so, it is essential to analyze lepton spectra
and perform ``meaningful'' Dalitz plot studies.
To quantify ``meaningful'', it is instructive to
compare to analyses on $K_{e4}$ decays.
With a sample size of 30,000 events which became available in 1977,
one was able to begin extracting dynamical information.
Precise measurements are now possible,
with NA48/2 and E685 each having accumulated
400,000 events~\cite{Pislak:2001bf,Batley:2004cp}.
For charm we are nowhere near that level:
CLEO-$c$ will have about 10,000 semileptonic charm decays --
comparable to kaon studies in the late 1970s.
Since for charm the phase space is larger,
thereby opening more domains of interest,
a reasonable target sample size is $10^6$ events,
which is far beyond the reach of CLEO-$c$, and most probably, of BESIII.

Three-family unitarity constraints on the CKM matrix yield
rather precise values for $|V_{cs}|$ and $|V_{cd}|$.
Using these, one can extract the form factors from analyses of
exclusive semileptonic charm decays.
Both the normalization and $q^2$ dependence must be measured.
Existing LQCD studies do not allow us to
determine the latter from first principles;
instead a parametrization originally proposed by Becirevic and Kaidalov ($BK$)
is used~\cite{Becirevic:1999kt}.
Recent and forthcoming results from CLEO-$c$, \babar\ and
\belle~\cite{AsnerCKM06, Widhalm:2006wz} are expected to be statistics limited,
and will not reach the desired $1$--$3\%$ level.

The current status can be characterized by comparing the measured value
of the ratio $R_{sl}$,
which is independent of $|V_{cd}|$,
to that inferred from a recent LQCD calculation~\cite{Aubin:2005ar}:
\begin{equation}
  R_{sl} = \sqrt{
    \frac{
      \Gamma(D^+ \rightarrow \mu^+ \nu_\mu)
    }{
      \Gamma(D \rightarrow \pi e \nu _e)
    }
  } =
  \left\{
    \begin{array}{c@{\hspace{3mm}}c}
      0.237 \pm 0.019 & ({\rm exp}) \\
      0.212 \pm 0.028 & ({\rm theo})\, . \\
    \end{array}
  \right.
\end{equation}
The values are nicely consistent,
yet both are still far from the required level of precision.


While operation in the $\Upsilon$ region will produce
large quantities of charm hadrons,
there are significant backgrounds and
one pays a price in statistics when using kinematic
constraints to infer neutrino momenta, \etc.
On the other hand, even a limited run at charm threshold
will generate the statistics required to
study (semi)leptonic decays with the desired accuracy.
Assuming that systematic uncertainties in tracking and muon identification
will provide a limit to the precision at the $0.5\%$ level,
we estimate the integrated luminosity from threshold running required
to achieve a similar statistical uncertainty.
As shown in Table~\ref{tab:statl} we expect to be able to measure
$f_{D^+}$, $f_{D_s}$ and their ratio with better than $0.5\%$
statistical uncertainty with integrated luminosities
of at least $100 \ {\rm fb}^{-1}$.
\begin{table}[!htbp]
    \caption{
      Statistics required to obtain $0.5\%$ statistical uncertainties
      on corresponding branching fractions using double-tagged events,
      when running at threshold.
    }
    \label{tab:statl}
  \begin{center}
    \begin{tabular}{cc}
      \hline \hline
      Channel & Integrated luminosity  \\
      & (${\rm fb}^{-1}$)  \\
      \hline
      $D^+ \rightarrow \mu^+ \nu_{\mu}$  & 500  \\
      $D_s^+ \rightarrow \mu^+ \nu_{\mu}$  & 100  \\
      \hline
    \end{tabular}
  \end{center}
\end{table}

For semileptonic decays, a case-by-case study is necessary.
One also has to distinguish between merely determining the branching ratio
and performing a ``meaningful'' Dalitz plot analysis, as discussed above.
The required integrated luminosities are given in Table~\ref{tab:statsl}.
It is clear that the $\sim 150 \ {\rm fb}^{-1}$ anticipated
from one month of running in the threshold region
would provide the desired statistics for most measurements.
Note that while $D_s$ mesons are not produced at the $\psi(3770)$,
short runs at other energies are possible.
\bigskip
\begin{table}[!b]
    \caption{
      Statistics required to obtain $0.5\%$ statistical uncertainties
      on corresponding branching fractions (column 2) or
      one million signal events (column 3) using double tagged events,
      when running at threshold.
    }
  \begin{center}
    \begin{tabular}{lcc}
      \hline \hline
      Channel & Integrated luminosity & Integrated luminosity \\
      & (${\rm fb}^{-1}$)  & (${\rm fb}^{-1}$)  \\
      \hline
      $D^0 \to K^- e^+ \nu_e$     & 1.3 &   33 \\
      $D^0 \to K^{*-} e^+ \nu_e$  & 17  &  425 \\
      $D^0 \to \pi^- e^+ \nu_e$   & 20  &  500 \\
      $D^0 \to \rho^- e^+ \nu_e$  & 45  & 1125 \\
      $D^+ \to \KS e^+ \nu_e$   &  9  &  225 \\
      & \\
      $D^+ \to \bar{K}^{*0} e^+ \nu_e$  &    9 &   225 \\
      $D^+ \to \pi^{0} e^+ \nu_e$       &   75 &  1900 \\
      $D^+ \to \rho^{0} e^+ \nu_e$      &  110 &  2750 \\
      & \\
      $D_s^+ \to \phi e^+ \nu_e$        &   85 &  2200 \\
      $D_s^+ \to \KS e^+ \nu_e$       & 1300 & 33000 \\
      $D_s^+ \to K^{*0} e^+ \nu_e$      & 1300 & 33000 \\
      \hline
    \end{tabular}
    \label{tab:statsl}
  \end{center}
\end{table}




\subsection{Precision CKM Measurements}
\label{CKMPREC}

Studies of leptonic decay constants and semileptonic form factors
will yield a set of measurements,
including $\left| V_{cd}\right|$ and $\left| V_{cs}\right|$,
at the few percent level.
These measurements will constrain theoretical calculations,
and those that survive will be validated for use in a variety of areas
in which interesting physics cannot be extracted without theoretical input.
This broader impact of charm measurements extends beyond
those measurements that can be performed directly at charm threshold, and
has a large impact on the precision determination of CKM matrix elements.

The determination of $\left| V_{td}\right|$ and $\left| V_{ts}\right|$
is limited by ignorance of $f_B\sqrt{B_{B_d}}$ and $f_{B_s}\sqrt{B_{B_s}}$;
improved determinations of $f_B$ and $f_{B_s}$ are required.
Precision measurements of $f_D$ and $f_{D_s}$ can validate
the theoretical treatment of the analogous quantities for $B$ mesons.
Similarly, improved form factor calculations in the decays
$D \to \pi \ell \nu$ and $D \to \rho \ell \nu$
and inclusive semileptonic charm decays will enable
improved precision in $\left|V_{ub}\right|$ and $\left|V_{cb}\right|$.

The precision measurement of the UT angle $\gamma$
depends on decays of $B$ mesons to final states
containing neutral $D$ mesons.
A variety of charm measurements impact these analyses, including:
improved constraints on charm mixing amplitudes,
 -- important for the GLW method~\cite{Gronau:1991dp,Gronau:1990ra},
measurements of relative rates and strong phases
between Cabibbo-favoured and -suppressed decays
 measurement of the relative rate and relative strong phase $\delta$
 between $D^0$ and $\overline D^0$ decay to $K^+\pi^-$ --
 important for ADS method\cite{Atwood:1996ci,Atwood:2000ck},
and studies of charm Dalitz plots tagged by hadronic flavor or
$\CP$ content~\cite{Giri:2003ty,Bondar:2005ki,White:2007br}.
Note that the latter two measurements
can only be performed with data from charm threshold.

\mysubsubsection{Overconstraining the Unitarity Triangle}

At present three-family unitarity constraints yield more precise values
for $|V_{cs}|$ and $|V_{cd}|$ than direct measurements.
Since it is conceivable that a fourth family exists
(with neutrinos so heavy that the $Z^0$ could not decay into them),
one would like to obtain more accurate direct determinations.
This should be possible if LQCD is indeed validated
at the ${\cal O}(1\%)$ level through its predictions
on form factors and their ratios.

From four-family unitarity,
and using current experimental constraints~\cite{Yao:2006px}
we can infer for a fourth quark doublet $(t^{\prime},b^{\prime})$:
\begin{eqnarray}
  |V_{cb'}| & = & \sqrt{1 - |V_{cd}|^2 - |V_{cs}|^2 - |V_{cb}|^2} \lsim 0.5~,\\
  |V_{t's}| & = & \sqrt{1 - |V_{us}|^2 - |V_{cs}|^2 - |V_{ts}|^2} \lsim 0.5~.
\end{eqnarray}
These loose bounds are largely due to the $10\%$ error on $|V_{cs}|$.


\subsection{Charm as a Window to New Physics}
\label{HYPOGEN}

While significant progress can be guaranteed for the
Standard Model studies outlined above,
the situation is much less certain concerning the search for New Physics.
No sign of it has yet been seen,
but we have only begun to approach the regime of experimental sensitivity
in which a signal for New Physics could realistically emerge in the data.
The interesting region of sensitivity extends
several orders of magnitude beyond the current status.

New Physics scenarios in general induce flavor-changing neutral currents
that {\em a priori} have no reason to be as strongly suppressed as in the Standard Model.
More specifically, they could be substantially stronger
for up-type than for down-type quarks;
this can occur in particular in models that
reduce strangeness-changing neutral currents
below phenomenologically acceptable levels through an alignment mechanism.

In such scenarios,
charm plays a unique role among the up-type quarks $u$, $c$ and $t$;
for only charm allows the full range of probes for New Physics.
Since top quarks do not hadronize~\cite{Bigi:1986jk},
there can be no $T^0\bar{T}^0$ oscillations
(recall that hadronization, while hard to bring under theoretical control,
enhances the observability of $\CP$ violation).
As far as $u$ quarks are concerned, $\pi^0$, $\eta$ and $\eta ^{\prime}$
do not oscillate, and decay electromagnetically, not weakly.
$\CP$ asymmetries are mostly ruled out by $\CPT$ invariance.
Our basic contention can then be formulated as follows:
charm transitions provide a unique portal for a novel access
to flavor dynamics with the experimental situation being {\em a priori}
quite favourable.
The aim is to go beyond ``merely'' establishing the existence of
New Physics around the TeV scale --
we want to identify the salient features of this New Physics as well.
This requires a comprehensive study,
\ie, that we also search in unconventional areas such as charm decays.

\mysubsubsection{On New Physics scenarios}
\label{CHARMNP}

In a scenario in which the LHC discovers direct evidence of SUSY
via observation of sleptons or squarks,
the \sff\ program becomes even more important.
The sfermion mass matrices are a new potential source of flavor mixing
and $\CP$ violation and contain information about the SUSY-breaking mechanism.
Direct measurements of the masses can only constrain
the diagonal elements of this matrix.
However, off-diagonal elements can be measured through the study
of loop-mediated heavy flavor processes.
As a specific example, a minimal flavor violation scenario such as mSUGRA
with moderate $\tan \beta$, could result in a SUSY partner mass spectrum
that is essentially indistinguishable from an SU(5) GUT model
with right-handed neutrinos.
However the mSUGRA scenario would be expected to yield no observable effects
in the heavy flavor sector,
whereas the SU(5) model is expected to produce measurable effects
in time-dependent $\CP$ violation in penguin-mediated hadronic
and radiative decays.

While there is no compelling scenario that would generate observable effects
in charm, but not in beauty and strange decays,
it is nevertheless reassuring that such scenarios do exist.
One should keep in mind that New Physics signals in charm $\CP$ asymmetries
are particularly clean,
since the Standard Model background (which often exists in $B$ decays) is largely absent.
The consequence is that New Physics could produce signals
that exceed Standard Model predictions by an order of magnitude or more --
something that is of great help in interpreting the signals.
We will focus on the most promising areas;
more details can be found in several
recent reviews~\cite{Burdman:2001tf,Bianco:2003vb,Burdman:2003rs}.

\mysubsubsection{$\DzDzb$ oscillations}
\label{DOSC}

Oscillations of neutral $D$ mesons driven by the two quantities
$x_D = \Delta M_D/\Gamma_D$ and $y_D = \Delta \Gamma_D/2\Gamma_D$
lead to an effective violation of the Standard Model
$\Delta C = \Delta Q$ and $\Delta C = \Delta S$
rules in semileptonic and nonleptonic channels.
The status of the Standard Model prediction can be summarized as~\cite{Bianco:2003vb}:
while one predicts $x_D \sim {\cal O}(10^{-3}) \sim y_D$,
at present one cannot rule out $x_D, \, y_D \sim 0.01$.

Many different charm decay modes can be used to search for charm mixing.
The appearance of ``wrong-sign'' kaons in semileptonic decays
would provide direct evidence for $\DzDzb$ oscillations
(or another process with origin beyond the Standard Model).
The wrong-sign hadronic decay $D^0 \to K^+\pi^-$
is sensitive to linear combinations of the mass and lifetime differences,
denoted $x_D^{\prime2}$ and $y_D^\prime$.
The relation of these parameters to $x_D$ and $y_D$ is controlled
by a strong phase difference.
Direct measurements of $x_D$ and $y_D$
independent of unknown strong interaction phases,
can also be made using time-independent studies of amplitudes present
in multi-body decays of the $\Dz$, for example, $\Dz\to\KS \pip\pim$.
Direct evidence for $y_D \ne 0$ can also appear through lifetime differences
between decays to $\CP$ eigenstates.
The measured quantity in this case, $y_{\CP}$, is equivalent to $y_D$
in the absence of $\CP$ violation.
Another approach is to study quantum correlations near
threshold~\cite{Bianco:2003vb,Bigi:1989ah,Asner:2005wf}
in $e^+e^- \to \Dz\Dzb (\pi^0)$ and in $e^+e^- \to \Dz \Dzb \gamma$,
which yield $C$-odd and $C$-even $\Dz\Dzb$ pairs, respectively.




Very recently, several new results have suggested that charm mixing may
be at the upper end of the range of Standard Model predictions.
\babar\ finds evidence for oscillations in $D^0 \to K^+\pi^-$
with $3.9\sigma$ significance~\cite{Aubert:2007wf},
while \belle\ sees a $3.2\sigma$ effect in $D^0 \to K^+K^-$,
with results using $D^0 \to \KS\pi^+\pi^-$
supporting the claim~\cite{Abe:2007dt}.
These results are consistent with previous measurements,
some of which had hinted at a mixing
effect~\cite{Godang:1999yd,Abe:2003ys,Zhang:2006dp,Aubert:2006kt,Asner:2005sz}.
The results are not systematics limited,
and further improvements are anticipated.

\begin{figure}[t]
  \begin{center}
    \includegraphics[angle=0, width=0.49\textwidth]{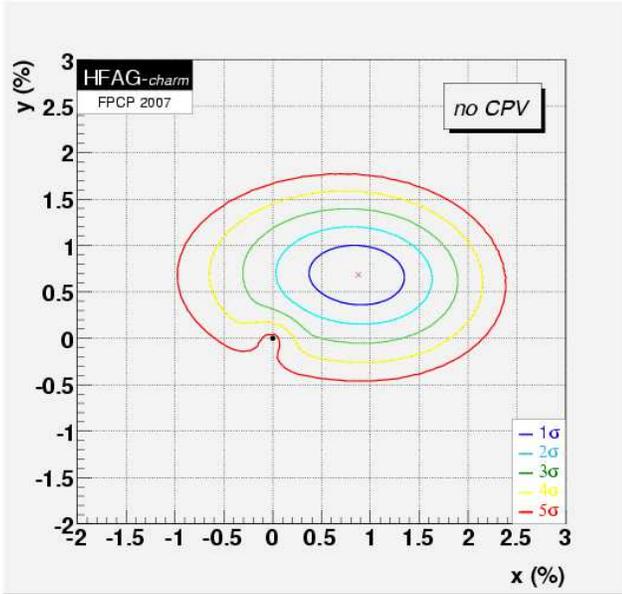}
    \caption{
      Likelihood contours in the $\left( x_D, y_D \right)$ plane
      from HFAG~\cite{hfag}.
      These preliminary world averages use all available charm mixing results.
    }
    \label{fig:HFAG_charm}
  \end{center}
\end{figure}

The charm decays subgroup of the Heavy Flavor Averaging Group~\cite{hfag}
is preparing world averages of all the charm mixing measurements,
taking into account correlations between the measured quantities.
A preliminary average is available, giving:
\begin{center}
  $x_D = \left( 8.7^{+3.0}_{-3.4} \right) \times 10^{-3}$
  \hspace{1mm} and \hspace{1mm}
  $y_D = \left( 6.6^{+2.1}_{-2.0} \right) \times 10^{-3}$\,.
\end{center}
Contours in the $\left( x_D, y_D \right)$ plane are shown
in Fig.~\ref{fig:HFAG_charm}.
The significance of the oscillation effect in
the preliminary world averages exceeds $5\sigma$.

At present no clear signal has emerged.
Since no single measurement exceeds $5\sigma$ significance,
it is too early to consider charm oscillations as definitively established.
Nonetheless, even if one accepts the central 
The interpretation of these new results in terms of New Physics is inconclusive.
For one thing, it is not yet clear whether the effect is caused by
$x_D \neq 0$ or $y_D \neq 0$ or both,
though the latter is favored and this point may be clarified soon.
As shown in Table~\ref{tab:charm},
\superb\ will be able to observe both lifetime and mass differences
in the $\Dz$ system, if they lie in the range of Standard Model predictions.
It should be noted that the full benefit of measurements
in the $\Dz \to K^+\pi^-$ system (and indeed for other hadronic decays)
can only be obtained if the strong phases are measured.
This can be achieved with a short ($\sim 1$ month) period
of data taking at charm threshold.

A serious limitation in the interpretation of charm oscillations
in terms of New Physics is the theoretical uncertainty on
the Standard Model prediction.
Nonetheless, if oscillations indeed occur at the level suggested
by the latest results, this will open the window to searches for
$\CP$ asymmetries that do provide unequivocal New Physics signals.

\begin{table}[htb]
  \caption{
    Summary of the expected precision on charm mixing parameters.
    For comparison we put the reach of the $B$~Factories at $2 \ {\rm ab}^{-1}$.
    The estimates for \superb\ assume that systematic uncertainties
    can be kept under control.
  }
  \label{tab:charm}
  \begin{center}
    \begin{tabular}{llcc}
      \hline \hline
      Mode &    & $B$~Factories & \superb\  \\
 & & ($2 \ {\rm ab}^{-1}$ & ($75 \ {\rm ab}^{-1}$
       \\
      \hline
      $\Dz \to K^+K^-$   & $y_{\CP}$       & $2$--$3 \times 10^{-3}$ & $5 \times 10^{-4}$ \\
      $\Dz \to K^+\pi^-$ & $y_D^\prime$    & $2$--$3 \times 10^{-3}$ & $7 \times 10^{-4}$ \\
                         & $x_D^{\prime2}$ & $1$--$2 \times 10^{-4}$ & $3 \times 10^{-5}$ \\
      $\Dz \to \KS\pi^+\pi^-$ & $y_D$      & $2$--$3 \times 10^{-3}$ & $5 \times 10^{-4}$ \\
                              & $x_D$      & $2$--$3 \times 10^{-3}$ & $5 \times 10^{-4}$ \\
      \hline
      Average            & $y_D$           & $1$--$2 \times 10^{-3}$ & $3 \times 10^{-4}$ \\
                         & $x_D$           & $2$--$3 \times 10^{-3}$ & $5 \times 10^{-4}$ \\
      \hline \hline
    \end{tabular}
  \end{center}
\end{table}


\mysubsubsection{$\CP$ Violation With and Without Oscillations}
\label{CPV}

Several factors favor dedicated searches for $\CP$ violation
in charm transitions:

\textbullet\   Within the Standard Model, the effective weak phase is highly diluted,
namely $\sim {\cal O}(\lambda ^4)$,
and can arise only in singly-Cabibbo-suppressed transitions,
where one expects asymmetries to reach the ${\cal O}(0.1 \%)$ level;
significantly larger values would signal New Physics.
Any asymmetry in Cabibbo-allowed or -doubly suppressed channels
requires the intervention of New Physics --
except for $D^{\pm}\to \KS\pi ^{\pm}$~\cite{Bianco:2003vb}
where the $\CP$ impurity in $\KS$ induces an asymmetry of $3.3 \times 10^{-3}$.
CLEO-c measures $A_{CP} = (-0.6 \pm 1.0 \pm 0.3)\%$\cite{Dobbs:2007zt}.
One should keep in mind that in going from Cabibbo-allowed to
Cabibbo-singly and -doubly suppressed channels,
the Standard Model rate is suppressed by factors of about
twenty and four hundred, respectively.
One would expect that this suppression will enhance
the visibility of New Physics.

\textbullet\   Strong phase shifts required for direct $\CP$ violation to emerge
are, in general, large, as are the branching ratios into relevant modes.
Although large final state interactions complicate the interpretation of
an observed signal in terms of the microscopic parameters
of the underlying dynamics, they enhance its observability.

\textbullet\   With the Standard Model providing one amplitude,
observable $\CP$ asymmetries can be linear in New Physics amplitudes --
unlike the case for rare decays -- thus increasing the sensitivity.

\textbullet\   Decays to multibody final states contain more dynamical information
than given by their widths;
their decay distributions as described by Dalitz plots or $T$-odd moments
can exhibit $\CP$ asymmetries that might be considerably larger
than those for the width.
Final state interactions,
while not necessary for the emergence of such effects,
can produce a signal that can be disentangled from New Physics effects
by comparing $T$-odd moments for $\CP$ conjugate modes~\cite{Link:2005th}.

\textbullet\   The distinctive channel $D^{*\pm} \to D \pi^{\pm}$ provides a powerful tag
on the flavor identity of the neutral $D$ meson.

The notable ``fly in the ointment'' in searching for $\CP$ violation in
the charm sector is that $\DzDzb$ oscillations are slow.
Nevertheless one should accept this challenge:
$\CP$ violation involving $\DzDzb$ oscillations
is a reliable probe of New Physics: the asymmetry is controlled by
$\sin(\Delta m_D t) \, \times \, \Im (q/p)\bar{\rho} (D \to f)$.
In the Standard Model both factors are small, namely $\sim {\cal O}(10^{-3})$,
making such an asymmetry unobservably tiny --
unless there is New Physics (see, \eg,~\cite{Grossman:2006jg,Agashe:2004cp}).
$\DzDzb$ oscillations, $\CP$ violation and New Physics
might thus be discovered simultaneously in a transition.
Such effects can be searched for in final states common to
$\Dz$ and $\Dzb$ such as $\CP$ eigenstates (\eg,~$\Dz \to K^+K^-$)
doubly Cabibbo suppressed modes (\eg,~$\Dz \to K^+\pi^-$)
or three-body final states (\eg\ $\Dz \to \KS\pi^+\pi^-$).
Undertaking time-dependent Dalitz plot studies\cite{Asner:2005sz,Abe:2007dt} requires a
high initial overhead,
yet in the long run this should pay handsome dividends,
since Dalitz plot analyses can invoke many internal correlations
that, in turn, serve to control systematic uncertainties.
Such analyses may allow the best sensitivity to New Physics.

\mysubsubsubsection{Direct $\CP$ violation}

 $\CP$ violation in $\Delta C = 1$ dynamics can be searched
 for by comparing partial widths for $\CP$ conjugate channels.
 For an observable effect two conditions have to be satisfied simultaneously:
 a transition must receive contributions from two coherent amplitudes with
 (a) different weak and (b) different strong phases.
 While condition (a) is just the requirement of $\CP$ violation
 in the underlying dynamics, condition (b) is needed
 to make the relative weak phase observable.
 Since the decays of charm hadrons proceed in the nearby presence
 of many hadronic resonances inducing virulent final state interactions (FSI),
 requirement (b) is in general easily met;
 thus it provides no drawback for the observability of a $\CP$ asymmetry --
 albeit it does for its microscopic interpretation.

 As already mentioned CKM dynamics does not support any $\CP$ violation
 in Cabibbo allowed and doubly suppressed channels
 due to the absence of a second weak amplitude.
 In singly Cabibbo suppressed transitions one expects $\CP$ asymmetries,
 albeit highly diluted ones of order
 $\lambda^4 \sim 10^{-3}$ or less~\cite{Grossman:2006jg}.


\mysubsubsubsection{$\CP$ asymmetries involving oscillations}

 For final states that are common to $D^0$ and $\bar D^0$ decays
 one can search for $\CP$ violation manifesting itself with the help of
 $D^0$--$\bar{D}^0$ oscillations in qualitative --
 though certainly not quantitative -- analogy to $B_d \to J/\psi \KS$.
 Such common states can be $\CP$ eigenstates --
 like $D^0 \to K^+K^-/\pi^+\pi^-/\KS\eta^{(\prime)}$ --, but do not have to be:
 two very promising candidates are $D^0 \to \KS \pi^+\pi^-$,
 where one can bring the full Dalitz plot machinery to bear,
 and $D^0 \to K^+\pi^-$ {\it vs.} $\bar D^0 \to K^-\pi^+$,
 since its Standard Model amplitude is doubly Cabibbo suppressed.
 Undertaking time-dependent Dalitz plot studies requires a
 higher initial overhead, yet in the long run this should pay handsome dividends
 exactly since Dalitz analyses can invoke many internal correlations
 that in turn serve to control systematic uncertainties.


\subsubsection{Experimental Status and Future Benchmarks}
\label{BENCH}

Time-integrated $\CP$ asymmetries have been searched for
and sensitivities of order $1\%$ [several $\%$]
have been achieved for Cabibbo-allowed and -singly suppressed modes with two
[three] body final states~\cite{Shipsey:2006gf}.
A Dalitz-plot analysis of time-integrated $\CP$ asymmetries provides constraints ${\cal O}(10^{-3})$\cite{Asner:2003uz}.
Time-dependent $\CP$ asymmetries
(\ie, those involving $\DzDzb$ oscillations)
are still largely {\it terra incognita}.

Since the primary goal is to establish the intervention of New Physics,
one ``merely'' needs a sensitivity level above the reach of the Standard Model;
``merely'' does not mean this can easily be achieved.
As far as direct $\CP$ violation is concerned,
this means asymmetries down to the $10^{-3}$ or $10^{-4}$ level in
Cabibbo-allowed channels
and down to the $1\%$ level or better in doubly Cabibbo-suppressed modes.
In Cabibbo-singly-suppressed decays one wants to reach the $10^{-3}$ range
(although CKM dynamics can produce effects of that order,
future advances might sharpen the Standard Model predictions).
For  time-dependent asymmetries in
$D^0 \to \KS\pi^+\pi^-$, $K^+K^-$, $\pi^+\pi^-$ \etc,
and in $D^0 \to K^+\pi^-$,
one should strive for the
${\cal O}(10^{-4})$ and ${\cal O}(10^{-3})$ levels, respectively.

When striving to measure asymmetries below the $1\%$ level,
one has to minimize systematic uncertainties.
There are at least three powerful weapons in this struggle:
i) resolving the time evolution of asymmetries that are controlled by
$x_D$ and $y_D$, which requires excellent vertex detectors;
ii) Dalitz plot consistency checks;
iii) quantum statistics constraints on distributions, $T$-odd moments,
\etc~\cite{Bigi:1989ah}.

\mysubsubsection{Experimental reach of New Physics searches}

In this section we briefly summarize the experimental reach
of \superb\ for New Physics sensitive channels in the charm sector.
Table~\ref{tab:charm_rare} shows the expected $90\%$ confidence level
upper limits that may be obtained on various important rare $D$ decays,
including suppressed flavor-changing neutral currents,
lepton flavor-violating and lepton number-violating channels,
from one month of running at the $\psi(3770)$.
It is expected that the results from running at the $\FourS$
will be systematics limited before reaching this precision.

For studies of $\DzDzb$ mixing,
running in the $\Upsilon$ region appears preferable,
and, if the true values of the mixing parameters are unobservably small,
the upper limits on both $x_D$ and $y_D$ can be driven to below $0.1\%$
in several channels ($D^0 \to K^+\pi^-$, $K^+K^-$, $\KS\pi^+\pi^-$, \etc)
Therefore, \superb\ can study charm
mixing if $x_D$ and $y_D$ lie within the ranges predicted by the Standard Model, and recently observed.
The sensitivity to mixing-induced $\CP$ violation effects
obviously depends strongly on the size of the mixing parameters.
If one or both of $x_D$ and $y_D$ are ${\cal O}(1\%)$,
as indicated by the most recent results,
\superb\ will be able to make stringent tests of
New Physics effects in this sector.

The situation for searches of direct $\CP$ violation is clearer:
the \superb\ statistics will be sufficient to observe the Standard Model effect
of $\sim 3 \times 10^{-3}$ in $D^+ \to \KS\pi^+$~\cite{Bianco:2003vb},
and other channels can be pursued to a similar level. 
Within three body modes, uncertainties in the Dalitz model are likely
to become the limiting factor.
However, model-independent $T$-odd moments can be constructed
in multibody channels,
and limits in the $10^{-4}$ region appear obtainable.

\begin{table}[!t]
    \caption{
      Expected $90\%$ confidence level upper limits that may be obtained on
      various important rare $D$ decays,
      from 1 month of \superb\ running at the $\psi(3770)$.
    }
    \label{tab:charm_rare}
  \begin{center}
    \begin{tabular}{lc}
      \hline \hline
      Channel & Sensitivity \\
      \hline
      $D^0 \to e^+e^-$,       $D^0 \to \mu^+\mu^-$      & $1 \times 10^{-8}$ \\
      $D^0 \to \pi^0 e^+e^-$, $D^0 \to \pi^0 \mu^+\mu^-$& $2 \times 10^{-8}$ \\
      $D^0 \to \eta  e^+e^-$, $D^0 \to \eta  \mu^+\mu^-$& $3 \times 10^{-8}$ \\
      $D^0 \to \KS   e^+e^-$, $D^0 \to \KS   \mu^+\mu^-$& $3 \times 10^{-8}$ \\
      $D^+ \to \pi^+ e^+e^-$, $D^+ \to \pi^+ \mu^+\mu^-$& $1 \times 10^{-8}$ \\
& \\
      $D^0 \to e^\pm\mu^\mp$        & $1 \times 10^{-8}$ \\
      $D^+ \to \pi^+ e^\pm\mu^\mp$  & $1 \times 10^{-8}$ \\
      $D^0 \to \pi^0 e^\pm\mu^\mp$  & $2 \times 10^{-8}$ \\
      $D^0 \to \eta  e^\pm\mu^\mp$  & $3 \times 10^{-8}$ \\
      $D^0 \to \KS   e^\pm\mu^\mp$  & $3 \times 10^{-8}$ \\
& \\
      $D^+ \to \pi^- e^+e^+$,       $D^+ \to K^-   e^+e^+$       & $1 \times 10^{-8}$ \\
      $D^+ \to \pi^- \mu^+\mu^+$,   $D^+ \to K^-   \mu^+\mu^+$   & $1 \times 10^{-8}$ \\
      $D^+ \to \pi^- e^\pm\mu^\mp$, $D^+ \to K^-   e^\pm\mu^\mp$ & $1 \times 10^{-8}$ \\
      \hline
    \end{tabular}
  \end{center}
\end{table}

\subsection{Summary: Charm Physics at Super$B$}
\label{SUMMCHARM}
One does not have to be an incorrigible optimist to argue
that the best might still be ahead of us in the exploration
of the weak decays of charm hadrons.
Detailed studies of leptonic and semileptonic charm decays will allow
experimental verification of improvements in lattice QCD calculations,
down to the required ${\cal O}(1\%)$ level of precision.
This will result in significant improvements in the precision of
CKM matrix elements.
The possibility to operate with $e^+e^-$ collision energies
in the charm threshold region further extends the physics reach
and the charm program of the \sff.

While no evidence for New Physics has yet been found in charm decays,
the searches have only recently entered a domain
where one could realistically hope for an effect.
New Physics typically induces flavor-changing neutral currents.
Those could be considerably less suppressed for
up-type than for down-type quarks.
Charm quarks are unique among up-type quarks in the sense that
only they allow to probe the full range of phenomena induced by
flavor changing neutral currents,
including $\CP$ asymmetries involving oscillations.

There is little Standard Model background to New Physics signals in charm $\CP$ asymmetries,
and what there is will probably be under good control
by the time \superb\ starts operating.
Baryogenesis -- necessary to explain the observed matter-antimatter asymmetry
in our Universe --
requires a new source of $\CP$ violation beyond that of the Standard Model.
Such new sources can be probed in charm decays on
three different Cabibbo levels, differing in rates
by close to three orders of magnitude.
With the Standard Model providing one amplitude, observable $\CP$ asymmetries
can be linear in a New Physics amplitude, thus
greatly enhancing their sensitivity.
Finally, as stated repeatedly,
the goal has to be to identify salient features of the anticipated New Physics
beyond ``merely'' ascertaining its existence.
This will require probing channels with
one or even two neutral mesons in the final state --
something that is possible only in an $e^+e^-$ production environment.
CLEO-$c$ and BESIII are unlikely to find $\CP$ asymmetries in charm decays,
and the $B$ Factory results will still be statistics limited.
A \sff\ would allow conclusive measurements.
\superb, with data taken at the $\FourS$ and near threshold, will complete the charm program down to the Standard Model level.






\def\B       {\ensuremath{B}\xspace}


\bibliography{sample}

\begin{thebibliography}{9}
\bibitem{ref:snowmassyear}
{The New Snowmass Year is an updating of the convention that multiplying peak luminosity by a ``year'' containing $10^7$ seconds provides a good measure of actual running time, the effects of accelerator and detector down time, dead time effects and the difference between peak and average luminosity. \pepii/\babar\ experience has shown that a ``New Snowmass Year'' with $1.5\times 10^7$ seconds is a more precise estimator of actual performance at a $B$ Factory.}
\end{thebibliography}


\end{document}